\pdfoutput=1

\documentclass[sigconf,preview]{aamas}

\usepackage{balance} 

\setcopyright{ifaamas}
\acmConference[AAMAS '25]{Proc.\@ of the 24th International Conference
on Autonomous Agents and Multiagent Systems (AAMAS 2025)}{May 19 -- 23, 2025}
{Detroit, Michigan, USA}{A.~El~Fallah~Seghrouchni, Y.~Vorobeychik, S.~Das, A.~Nowe (eds.)}
\copyrightyear{2025}
\acmYear{2025}
\acmDOI{}
\acmPrice{}
\acmISBN{}

\acmSubmissionID{1336}

\usepackage[utf8]{inputenc} 
\usepackage[T1]{fontenc}    
\usepackage{hyperref}       
\usepackage{url}            
\usepackage{booktabs}       
\usepackage{amsfonts}       
\usepackage{nicefrac}       
\usepackage{bookmark}
\usepackage{microtype}      
\usepackage{xcolor}         
\usepackage{graphicx}       
\usepackage{subcaption} 
\usepackage{algorithm}
\usepackage{algpseudocode}
\usepackage{enumitem}
\usepackage{amsthm}
\usepackage{amsmath}
\usepackage{nicefrac}       

\usepackage{amssymb}
\usepackage{multicol}
\usepackage{caption}
\usepackage{wrapfig}
\usepackage{pdfpages}
\usepackage{lipsum}

\usepackage{pgfplots}
\pgfplotsset{compat=1.18}
\usepackage{balance}

\usepackage[utf8]{inputenc}
\DeclareUnicodeCharacter{03B1}{\ensuremath{\alpha}}

\title[AAMAS-2025 Formatting Instructions]{Conflux-PSRO: Effectively Leveraging Collective Advantages in Policy Space Response Oracles}

\author{Yucong Huang}
\authornote{Both authors contributed equally to this research.}
\affiliation{
  \institution{School of Software and Microelectronics, Peking University}
  \city{Beijing}
  \country{China}
}

\author{Jiesong Lian}
\authornotemark[1]
\affiliation{
  \institution{Huazhong University of Science \& Technology}
  \city{Wuhan}
  \country{China}
}

\author{Mingzhi Wang}
\affiliation{
  \institution{Institute for Artificial Intelligence, Peking University}
  \city{Beijing}
  \country{China}
}

\author{Chengdong Ma}
\affiliation{
  \institution{Institute for Artificial Intelligence, Peking University}
  \city{Beijing}
  \country{China}
}
\author{Ying Wen}
\affiliation{
  \institution{Shanghai Jiao Tong University}
  \city{Shanghai}
  \country{China}
}
\email{ying.wen@sjtu.edu.cn}

\begin{abstract}
Policy Space Response Oracle (PSRO) with policy population construction has been demonstrated as an effective method for approximating Nash Equilibrium (NE) in zero-sum games. 
Existing studies have attempted to improve diversity in policy space, primarily by incorporating diversity regularization into the Best Response (BR). However, these methods cause the BR to deviate from maximizing rewards, easily resulting in a population that favors diversity over performance, even when diversity is not always necessary. Consequently, exploitability is difficult to reduce until policies are fully explored, especially in complex games.
In this paper, we propose \textbf{Conflux-PSRO}, which fully exploits the diversity of the population by adaptively selecting and training policies at \emph{state-level}. 
Specifically, Conflux-PSRO identifies useful policies from the existing population and employs a routing policy to select the most appropriate policies at each decision point, while simultaneously training them to enhance their effectiveness. Compared to the single-policy BR of traditional PSRO and its diversity-improved variants, the BR generated by Conflux-PSRO not only leverages the specialized expertise of diverse policies but also synergistically enhances overall performance.  
Our experiments on various environments demonstrate that Conflux-PSRO significantly improves the utility of BRs and reduces exploitability compared to existing methods.

\end{abstract}


\newcommand{\BibTeX}{\rm B\kern-.05em{\sc i\kern-.025em b}\kern-.08em\TeX}

\begin{document}


\pagestyle{fancy}
\fancyhead{}


\maketitle 


\section{Introduction}
\begin{figure}[ht!]
\centering          
\includegraphics[width=1\linewidth]{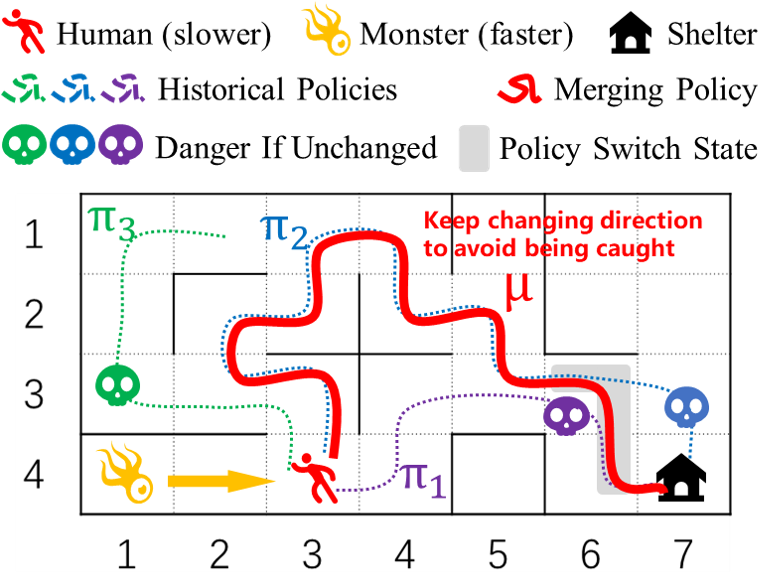}   
\caption{\textbf{Maze Game:} A 7x4 grid-based, two-player, zero-sum game between a \emph{human} (red) trying to reach the \emph{shelter} (black) and a \emph{monster} (yellow) attempting to capture the human. The human moves one grid per step, while the monster can move up to two in the same direction. The human explores three policies: $\pi_1$ (purple), $\pi_2$ (blue), and $\pi_3$ (green). $\pi_1$ and $\pi_2$ are more effective than $\pi_3$, but each leads to eventual danger if followed strictly. By switching between useful policies—starting with $\pi_2$ and switching to $\pi_1$ at critical points—the human can win, demonstrating how combining the strengths of historical policies creates a more robust policy $\pi_*$ (red), accelerating exploration and policy improvement.}
\label{fig:moe1}
\end{figure}
Most zero-sum games demonstrate strong non-transitivity~\cite{Czarnecki2020RealWG}, particularly in complex environments such as StarCraft~\cite{vinyals2019grandmaster,peng2017multiagent} and DOTA2~\cite{ye2020fullmoba}, which present significant challenges in both game theory and artificial intelligence. Addressing these challenges typically involves developing a set of policies that approximate a Nash Equilibrium (NE), enabling robust competition against any opponent. 
For larger-scale problems with complex game landscapes, the Policy Space Response Oracles (PSRO) algorithm provides an effective open-ended learning paradigm~\cite{lanctot2017unified}, iteratively approximating the NE by learning the best response (BR) to the opponent's mixed policies at each iteration.
To improve policy diversity, methods such as Diverse PSRO~\cite{nieves2021modelling}, BD\&RD-PSRO~\cite{liu2021unifying}, UDM-PSRO~\cite{liu2022unified}, and PSD-PSRO~\cite{yao2023policy} have been developed, expanding the policy hull and increasing generalization against a broader range of opponents. 

Despite these contributions, existing diversity-impoved PSRO variants primarily incorporate diversity regularization into the BR, easily causing the BR to deviate from its goal of maximizing rewards. 
As a result, these methods tend to prioritize diversity over performance, often blindly expanding the policy set to cover as many policies as possible.  Consequently, exploitability is difficult to reduce until policies are fully explored, which is both costly and challenging to achieve in complex games. On the contrary, sometimes the population requires the expanding of more adaptable policies within the NE support. 
As illustrated in Fig.~\ref{fig:moe1} (The Maze Game), this is a grid-based two-player zero-sum game played on a 7x4 grid, where $S(m,n)$ denotes the position at row $m$ and column $n$. There are two roles: the \emph{human} (red) and the \emph{monster} (yellow). The human moves first and aims to reach the \emph{shelter} (black) before being caught, while the monster's goal is to capture the human before they reach the shelter. The monster has a superior ability, moving either one or two steps in the same direction per time step, while the human can only move one step per time step. Now, assume we play as the human. Supposing the monster follows a fixed policy, chasing the human as closely as possible rather than intercepting. Thus, a reasonable idea for the human is to change direction frequently to avoid being quickly caught. Assume diverse-improved PSRO methods have explored three policies, represented by $\pi_1$ (purple), $\pi_2$ (blue), and $\pi_3$ (green). It is evident that $\pi_1$ and $\pi_2$ are stronger to $\pi_3$, as they both allow the human to continually change direction, preventing the monster from catching up quickly. However, none of these policies are perfect. If strictly followed, each will eventually lead the human to be caught at the positions marked as "Danger If Unchanged" due to consecutive moves in the same direction. 

One key insight is to focus on identifying and selecting relatively useful policies, leveraging them to construct more adaptable BR. 
In traditional diversity-improved PSRO methods, due to inherent deviation and the challenge of identifying useful policies, many weak policies, such as $\pi_3$ or many other potential policies that not marked in the Fig.~\ref{fig:moe1}, may be explored until all policies are covered. However, if we can recognize that $\pi_1$ and $\pi_2$ are relatively useful and combine their strengths—by selecting the $\pi_2$ at the starting point $S(4,3)$ and switching to $\pi_1$ policy at $S(3,6)$, combining into $\pi^*$ (red) —victory can be achieved. This shows that such a fine-grained combination of policies enables selecting the best historical policy at each state, outperforming any individual or mixed historical policies. This fine-grained combination accelerates exploration efficiency and yields stronger policies.

In this paper, we propose \textbf{Conflux-PSRO}, a novel method that converges the capabilities of a diverse population at \emph{state-level} by adaptively selecting policies for each state. 
Specifically, Conflux-PSRO identifies relatively useful historical policies as sub-policies (such as $\pi_1$ and $\pi_2$ in Fig.~\ref{fig:moe1}) and employs a routing policy to select the most appropriate sub-policy at each decision point (such as selecting $\pi_1$ at the state of $S(3,6)$ in Fig.~\ref{fig:moe1}) , ensuring that the decision-making process leverages the full potential of the population. Compared to the single-policy BR of traditional PSRO and its diversity-improved variants, the BR generated by Conflux-PSRO not only leverages the specialized expertise of diverse policies but also synergistically enhances overall performance. Through the accumulation of advantages over multiple steps within an episode, this fine-grained policy combination can achieve improved performance.
This conflex operation allows Conflux-PSRO to maintain a high level of flexibility and adaptability across various game scenarios over multiple states.
To ensure the population benefits from the routing policy, we distill the routing policy along with its sub-policies into a single policy, which is iteratively added to the population as a new BR. The process maintains the PSRO structure with singular policies and simplifies deployment and decision-making. We evaluate Conflux-PSRO on hard games. The experimental results demonstrate that Conflux-PSRO significantly improves the utility of BRs and reduces exploitability as compared to existing state-of-the-art methods, highlighting its effectiveness in generating robust and adaptable policies.

\section{Related Work}
\label{section-relatedwork}

\textbf{Ensemble Learning} architectures have been widely applied in reinforcement learning (RL) to enhance policy robustness, stability, and performance by leveraging multiple models or policies. Ensemble methods such as Bootstrapped DQN~\cite{osband2016deep} and Averaged-DQN~\cite{anschel2017averaged} combine multiple policies or value functions to mitigate overfitting, improve exploration, and provide more reliable uncertainty estimates. Recent approaches, including determinantal point processes and stochastic ensemble value expansion, emphasize maximizing policy diversity and enhancing exploration~\cite{sheikh2022dns, sharma2022deepevap, buckman2018sample}. Other research focuses on improving exploration through the use of policy or value function ensembles~\cite{jiang2017contextual, agarwal2020pc}. In the context of PSRO, leveraging policy ensembles to boost generalization shows great potential, but applying them directly to adjust to changing opponent strategies can lead to significantly higher computational costs~\cite{song2023ensemble}.

\textbf{Mixture of Experts (MoE)} models expand capacity by routing inputs to specialized expert networks through a gating mechanism~\cite{jacobs1991adaptive, shazeer2017outrageously}, enabling efficient scaling while maintaining computational feasibility~\cite{smith2021probMoE-RL}. The MoE framework has proven particularly advantageous in managing high-dimensional state spaces and enhancing policy representations, as demonstrated in reinforcement learning dialogue systems~\cite{lee2022MoE-RL-dialogue} and various other machine learning applications~\cite{doe2021survey-MoE-RL}. 
In this paper, we primarily use a routing-policy to model the gating mechanism of MoE, as detailed later. We integrate this routing-policy into PSRO for extensive-form games, leading to iterative improvements in population-based algorithms and better performance across policies. 

\textbf{The Policy Space Response Oracle (PSRO) }framework ~\cite{lanctot2017unified} generalizes the Double Oracle algorithm to continuous and high-dimensional policy spaces, iteratively building a policy population by computing BRs to the current meta-strategy to approximate a Nash equilibrium. Several methods have been introduced in PSRO to foster diversity within the policy space, mirroring techniques in RL designed to maximize exploration efficiency~\cite{sheikh2022dns, sharma2022deepevap, buckman2018sample}. Extensions like Diverse-PSRO ~\cite{balduzzi2019diverse}, BD\&RD-PSRO ~\cite{nieves2021modelling}, and PSD-PSRO ~\cite{yao2023policy} have introduced methods to enhance policy diversity within the population, aiming to better explore the policy space and reduce exploitability. These approaches focus on enlarging the gamescape or the Policy Hull to approximate a Nash equilibrium more closely. Our approach differs by focusing on training better BRs by converging powerful and diverse historical policies, rather than solely increasing policy diversity. 

\section{Notations and Preliminary}

\textbf{ Markov Decision Process (MDP) and Extensive-Form Games.} We consider a finite MDP defined by a state space \(S\), action space \(A\), transition probabilities \(P(s' | s, a)\), and a reward function \(R(s, a, s')\), where an agent interacts with the environment by choosing actions, transitioning to new states, and receiving rewards with the goal of maximizing cumulative rewards over time. Extending this to multi-agent scenarios, we analyze extensive-form games with perfect recall~\cite{hansen2004dynamic}, where players take sequential actions associated with world states \(w \in \mathcal{W}\). In an \(N\)-player game, the joint action space is \(\mathcal{A} = \mathcal{A}_1 \times \cdots \times \mathcal{A}_N\). For player \(i\), the set of legal actions at state \(w\) is \(\mathcal{A}_i(w)\), and the transition function \(\mathcal{T}(w, a)\) defines the probability of moving to the next world state \(w'\). Upon transitioning, player \(i\) observes \(o_i = \mathcal{O}_i(w, a, w')\) and receives a reward \(\mathcal{R}_i(w)\). The game concludes after a finite sequence of actions when a terminal state is reached.

\textbf{Game Metrics and Equilibrium.} An information set \( s_i \) for player \( i \) comprises the player's observations and actions up to that point: \( s_i(h) = (a_i^0, o_i^1, a_i^1, \ldots, o_i^t) \).  A history $h = (w^0, a^0, w^1, a^1, \ldots, w^t)$ is a sequence of states and actions starting from the initial state \( w^0 \). A policy \( \pi_i \) for player \( i \) maps each information set \( s_i \) to a probability distribution over actions. A strategy profile \( \pi = (\pi_1, \ldots, \pi_N) \) represents the policies of all players, with \( \pi_{-i} \) denoting the policies of all players except \( i \). When a policy \( \pi_i \) is learned via reinforcement learning, it is referred to as a policy. The expected value (EV) \( v_i^{\pi}(h) \) for player \( i \) is the expected sum of future rewards starting from history \( h \) under strategy profile \( \pi \). The EV for an information set \( s_i \) is \( v_i^{\pi}(s_i) \), and for the entire game, it is \( v_i(\pi) \). In two-player zero-sum games, the sum of the expected values satisfies \( v_1(\pi) + v_2(\pi) = 0 \) for all strategy profiles \( \pi \). A Nash equilibrium (NE) is a strategy profile \( \pi^* \) where no player can increase their EV by unilaterally deviating: \( v_i(\pi^*) = \max_{\pi_i} v_i(\pi_i, \pi^*_{-i}) \) for each player \( i \). The exploitability \( e(\pi) \) of a strategy profile \( \pi \) is defined as \( e(\pi) = \sum_{i \in \mathcal{N}} \max_{\pi'_i} v_i(\pi'_i, \pi_{-i}) \). A Best Response (BR) policy for player \( i \), denoted \( \mathbb{BR}_i(\pi_{-i}) \), maximizes their EV against \( \pi_{-i} \): $\mathbb{BR}_i(\pi_{-i}) = \arg\max_{\pi_i} v_i(\pi_i, \pi_{-i})$. An \( \epsilon \)-BR policy \( \mathbb{BR}^\epsilon_i(\pi_{-i}) \) for player \( i \) satisfies: $v_i(\mathbb{BR}^\epsilon_i(\pi_{-i}), \pi_{-i}) \ge v_i(\mathbb{BR}_i(\pi_{-i}), \pi_{-i}) - \epsilon$. An \( \epsilon \)-Nash equilibrium (\( \epsilon \)-NE) is a strategy profile \( \pi \) where each player's policy is an \( \epsilon \)-BR to the others: \( \pi \) is an \( \epsilon \)-NE if \( \pi_i \) is an \( \epsilon \)-BR to \( \pi_{-i} \) for each player \( i \). 

\textbf{Normal-Form Game and Restricted Game.} A \emph{normal-form game} is a single-step extensive-form game. An extensive-form game induces a normal-form game where the legal actions for player $i$ are its deterministic strategies $X_{s_i \in \mathcal{I}_i} \mathcal{A}_i(s_i)$. These deterministic strategies are called pure strategies. A mixed strategy is a distribution over a player's pure strategies. A \textit{restricted game} in the context of PSRO represents a modified form of a \emph{normal-form game} in which each player's strategy set is constrained to specific populations $\Pi^t_i$. During each iteration, the Meta-NE $\sigma$ is calculated based on these limited populations. Following this, each player augments their population by computing and integrating the BRs to the current equilibrium.

\section{Conflux-PSRO}
\label{section-method}
In this section, we develop a new BR-enhanced PSRO viariant, Conflux-PSRO. 
Unlike diversity-enhanced PSRO variants ~\cite{balduzzi2019diverse,nieves2021modelling,yao2023policy} that blindly increase diversity, Conflux-PSRO aims to identify useful historical policies and use them to construct stronger BRs, thereby achieving efficient exploration of the policy space. Specifically, Conflux-PSRO involves converge these useful policies into a more fine-grained combination, an operation we refer to as \textit{conflux}.

\subsection{The Conflux Operation }
\begin{figure}
\centering          
\includegraphics[width=1\linewidth]{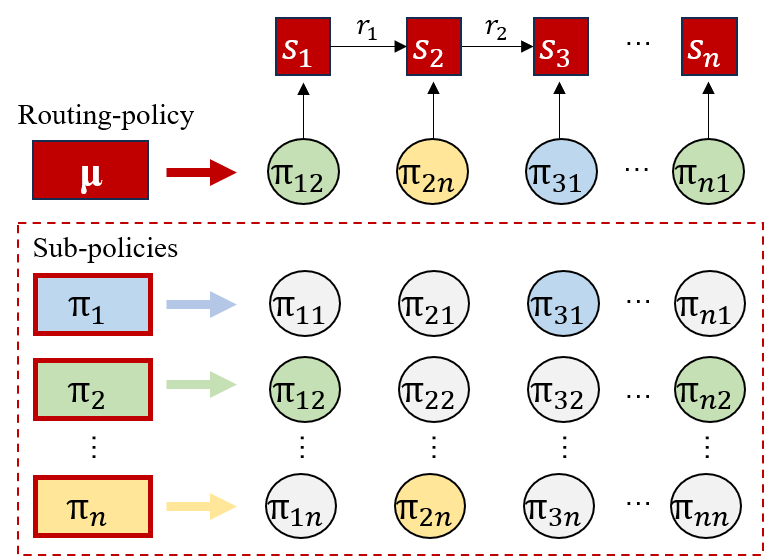}   
\caption{Construction of Routing-policy: A routing-policy serves as a decision-making layer that adaptively selects the most appropriate sub-policy for any given state, thereby optimizing the overall system's performance.}
\label{fig:moe2} 
\end{figure}

The \textbf{conflux} operation is designed to converge the capabilities of multiple sub-policies at \emph{state-level}, using an overarching \textbf{routing-policy} as shown in Fig.~\ref{fig:moe2}. The routing-policy
serves as a decision-making layer that adaptively selects the most appropriate sub-policy \emph{for any given state}, thereby optimizing the overall system's performance. We define a routing-policy \(\mu\) maps states \(s\) to one policy of sub-policies \(\{\pi_1, \pi_2, \ldots, \pi_n\}\). Specifically, \(\mu(i|s)\) denotes the probability of selecting sub-policy \(i\) (i.e., \(\pi_i\)) when in state \(s\). In this setting, the action-value function \(Q^\mu(s, i)\) under the routing-policy \(\mu\) is defined as:

\begin{equation}
Q^\mu(s, i) = \mathbb{E}_{\pi_i}\left[\sum_{t=0}^{\infty} \gamma^t R(s_t, a_t, s_{t+1}) \,\bigg|\, s_0 = s\right],
\label{}
\end{equation}

where \(\pi_i\) is the \(i\)-th sub-policy, and the expectation is over the trajectories generated by \(\pi_i\) starting from state \(s\). The Bellman equation for the action-value function in the context of the routing-policy is:

\begin{equation}
Q^\mu(s, i) = \sum_{a \in A} \pi_i(a | s) \sum_{s' \in S} P(s' | s, a) \left[ R(s, a, s') + \gamma \sum_{j=1}^{n} \mu(j | s') Q^\mu(s', j) \right].
\end{equation}

Our goal is to find an optimal routing-policy \(\mu^*\) that maximizes the state-value function \(V^\mu(s)\) for all states \(s \in S\). Formally, the optimization problem is:
\begin{equation}
\mu^*(s) = \arg\max_{\mu} V^\mu(s) = \arg\max_{\mu} \sum_{i=1}^{n} \mu(i | s) Q^\mu(s, i), \quad \forall s \in S.
\label{eq:bellman}
\end{equation}
An insight here is that optimizing the routing-policy \(\mu^*\) at the state level allows it to adaptively select the most suitable policy \(\pi_i\) from the population \(\{\pi_i\}\), potentially outperforming the meta-NE policy \(\sigma^{\text{NE}}\) optimized at the episode level. As mixed policies often outperform pure policies in game theory \cite{vonNeumann1944}, it follows that \(\mu^*\) could perform at least as well as \(\sigma^{\text{NE}}\) and any individual policy \(\pi_i\), i.e., \(\mu^* \geq \sigma^{\text{NE}} \geq \pi_i\). Though this is an intuitive consideration, it suggests that the BR generated via routing-policy could be stronger than those produced by traditional PSRO methods and their variants.
With this standing, the optimization of \(\mu^*\) can be achieved through various methods, including policy gradient techniques or other advanced algorithms.
To optimize within the PSRO framework (or other population training methods), we can leverage the population generated by PSRO for sub-policy selection and routing-policy training. 
The following sections will outline the steps shown in Fig.~\ref{fig:moe2}, i.e.,  identify useful policies and select them as sub-policies, namely \textbf{Sub-policies Selection} and  converge the selected sub-policies by learning a routing-policy, called \textbf{Routing-Policy Training}.

\subsection{Sub-policies Selection}
Selecting an appropriate number of sub-policies is critical in the design of routing policies. An excessive number of sub-policies can lead to increased computational complexity and a higher risk of overfitting, while too few may compromise both diversity and adaptability. Ensuring diversity and the strength of individual policies is essential. Although diversity helps reduce the risk of exploitation by unseen opponents, weak sub-policies can undermine the overall performance of the routing policy. Therefore, both diversity and strong policies are crucial~\cite{jacobs1991adaptive, shazeer2017outrageously}. \textbf{To identify powerful sub-policies}, we use the meta-NE to select policies with higher probabilities, forming an candidate set. Next, \textbf{to identify diverse sub-policies}, we apply a diversity metric~\cite{yao2023policy}, maximizing the distance between a candidate and the existing policy hull using the Kullback-Leibler divergence:

\vspace{-1em}
\begin{eqnarray}
\text{Diversity}(\pi, \mathcal{PH}) = \min_{\pi^{\prime}_ \in \mathcal{PH}}  \text{dist}(\pi  ,\pi^{\prime}) = \min_{\pi^{\prime}_ \in \mathcal{PH}}  D_\text{KL}(\pi  ||\pi^{\prime})
\label{selection}
\end{eqnarray}
As illustrated in in Algorithm \ref{alg:policy_selection}, to integrate both \textbf{powerful} and \textbf{diverse} identification methods. We first select the top $m$ policies with the highest Meta-NE probabilities. 
The top policy is added to the \textbf{selected pool} $\Pi_i^{\text{sele}}$, 
while the remaining $m-1$ policies are placed into the \textbf{candidate pool} $\Pi_i^{\text{cand}}$. 
Subsequently, using the defined diversity metric, we iteratively select one of candidate policies from the candidate pool that maximize the distance to the policy hull (PH) of the selected pool and add them to the selected pool. This process continues until the number of sub-policies reaches $n$. This two-layer selection ensures that the chosen sub-policies are not only powerful but also diverse, encompassing a wide range of capabilities and tactics to better adapt to different opponents.

\begin{algorithm}[t!]
\caption{Policy Selection}
\begin{algorithmic}[1] 
\State \textbf{Input:} Population of policies \( \Pi_i \), Meta-NE probabilities \( \sigma^t_i \), selection parameters \( m \) and \( n \).
\State \textbf{Output:} Selected sub-policies \( \Pi_i^{\text{sele}} \)
\State Sort \( \Pi_i \) in descending order based on Meta-NE probabilities.
\State \( \Pi_i^{\text{sele}} \leftarrow \{\pi_1\} \)
\State \( \Pi_i^{\text{cand}} \leftarrow \{\pi_2, \pi_3, \ldots, \pi_m\} \) 
\State Sample states \( S \) from buffer of \( \pi_1 \), \( a_1 \leftarrow  \pi_1(S), a_2 \leftarrow \pi_2(S), \ldots, a_m \leftarrow \pi_m(S) \)
\For{each pair \( (\pi_j, \pi_k) \) where \( j, k \in \{1, 2, \ldots, m\} \) and \( j \neq k \)}
    \State \( \text{dist}(\pi_j , \pi_k) \leftarrow D_\text{KL}(a_j || a_k) \)
\EndFor
\While{\(|\Pi_i^{\text{sele}}| < n\)}
    \State \( \pi^* \leftarrow \arg\max_{\pi \in \Pi_i^{\text{cand}}} \text{Diversity}(\pi, \mathcal{PH}(\Pi_i^{\text{sele}})) \) \Comment{Eq.~\eqref{selection}}
    \State \( \Pi_i^{\text{sele}} \leftarrow \Pi_i^{\text{sele}} \cup \{\pi^*\} \) 
    \State \( \Pi_i^{\text{cand}} \leftarrow \Pi_i^{\text{cand}} \setminus \{\pi^*\} \) 
\EndWhile
\State \textbf{Return} \( \Pi_i^{\text{sele}} \)
\end{algorithmic}
\label{alg:policy_selection}
\end{algorithm}

\subsection{Routing-Policy Training }
After selecting a limited number of sub-policies that are both powerful and diverse, we can optimize the performance of the routing-policy. In this paper, we utilize the Deep Q-learning Network (DQN) method~\cite{mnih2015human} for optimization. The objective function for DQN in the context of routing-policy is:
\begin{equation}
\mathcal{L}(\theta, \phi) = \mathbb{E} \left[ \left( r + \gamma \max_{j} Q_{\theta^-, \phi}(s', j) - Q_{\theta, \phi}(s, i) \right)^2 \right],
\end{equation}
where \(Q_{\theta, \phi}(s, i)\) represents the estimated action-value function that depends on both the routing-policy parameters \(\theta\) and the sub-policy parameters \(\phi\), which share the same structure and parameters, and \(\theta^-\) denote the parameters of the target networks for the routing-policy. The DQN algorithm optimizes this objective function by updating the network parameters through gradient descent:
\begin{align}
\label{eq:moe_train}
\theta &\leftarrow \theta - \alpha \nabla_\theta \mathcal{L}(\theta, \phi), \\
\phi &\leftarrow \phi - \beta \nabla_\phi \mathcal{L}(\theta, \phi),
\end{align}
where \(\alpha\) is the learning rate for the routing-policy, and \(\beta\) is the learning rate for the sub-policies. 
By including the sub-policy parameters \(\phi\) in the action-value function \(Q_{\theta, \phi}(s, i)\), we explicitly model the dependence of the Q-values on both the routing-policy and the sub-policies. This reflects the fact that any adjustments in the sub-policies will influence the expected returns, and thus the Q-values.
Given that DQN is an off-policy method and operates in a finite state and action (sub-policy) space, it is expected to converge to the optimal action-value function \(Q^*(s, i)\) under certain conditions. This enables us to derive the optimal routing-policy \(\mu^*\) by selecting the sub-policy with the highest Q-value in each state:
\begin{equation}
\mu^*(s) = \arg\max_{i} Q^*(s, i).
\end{equation}

By jointly training the routing-policy and sub-policies, we ensure that the entire architecture adapts cohesively. 

To ensure the population benefits from the routing-policy, we distill the routing-policy along with its sub-policies into a single policy \(\pi_{\text{distill}}\). Subsequently, we iteratively add \(\pi_{\text{distill}}\) to the population as a BR, thereby maintaining the PSRO structure with singular policies and simplifying deployment and decision-making. In the distillation training process, we incorporate three objectives: reward maximization, imitation of the routing-policy, and diversity promotion (proposed by \cite{yao2023policy}). By balancing these objectives, the distilled policy \(\pi_{\text{distill}}\) effectively  inherits the strengths of the routing-policy while effectively balancing exploitation (via reward maximization) and exploration (via diversity promotion). To optimize this objective, we compute the gradient of the loss function with respect to the parameters \(\theta\) of \(\pi_{\text{distill}}\):


\begin{equation}
\nabla_{\theta} \mathcal{L}_{\text{distill}} = \nabla_{\theta} \mathcal{L}_{\text{env}} + \nabla_{\theta} \mathcal{L}_{\text{imitation}} + \nabla_{\theta} \mathcal{L}_{\text{diversity}},
\label{eq:loss0}
\end{equation}


where:
\[
\nabla_{\theta} \mathcal{L}_{\text{env}} = \mathbb{E}_{(s, a) \sim \mathcal{D}} \left[ \nabla_{\theta} r(s, a) \right],
\label{eq:loss1}
\]
\[
\nabla_{\theta} \mathcal{L}_{\text{imitation}} = - \lambda_1 \, \mathbb{E}_{(s, a) \sim \mathcal{D}} \left[ \nabla_{\theta} D_{\text{KL}} \left( \pi_{\text{distill}}(a|s) \,\|\, \mu(a|s) \right) \right],
\label{eq:loss2}
\]
\[
\nabla_{\theta} \mathcal{L}_{\text{diversity}} = \lambda_2 \, \mathbb{E}_{s \sim \mathcal{D}} \left[ \nabla_{\theta} \min_{\pi^k \in H(\Pi^t)} D_{\text{KL}} \left( \pi_{\text{distill}}(\cdot|s) \,\|\, \pi^k(\cdot|s) \right) \right].
\label{eq:loss3}
\]



In the subsequent experimental section, we demonstrate that distillation does not significantly degrade population performance. The detailed algorithm for the distillation process can be found in the Appendix.

\subsection{Conflux-PSRO Algorithm}
After implementing the sub-policies selection and routing-policy training in \textit{conflux}, we can now construct the complete process of the Conflux-PSRO algorithm. By iteratively executing \textit{conflux}, we generate improved BRs, thereby enhancing the PSRO framework and its diversity-promoting variants. The pseudocode for the Conflux-PSRO algorithm is presented in Algorithm~\ref{xxx_psro}.

\begin{algorithm}[h]
\caption{Conflux-PSRO}
\label{xxx_psro}
\begin{algorithmic}[1]
\State \textbf{Input:} Initial policy sets for all players \(\Pi\)
\State Compute utilities \(U^{\Pi}\) for each joint policy \(\pi \in \Pi\)
\State Initialize meta-NE \(\sigma_i = \text{UNIFORM}(\Pi_i)\) for each player \(i\)
\For{\emph{e} = 1, 2, \dots}
    \For{\emph{player} \(i \in \{1, 2, \dots, n\}\)}
        \If {\textbf{conflux should run}}
            \State Select sub-policies \(\pi_i^{\text{selected}}\) via Algo. \ref{alg:policy_selection}
            \State Initialize routing-policy \(\mu_i\) over copied \(\pi_i^{\text{selected}}\)
            \State \textbf{Cross-Play} training for E episodes each:
            \State \quad Train \(\pi_{-i}^{\prime}\) over trajectory \(\rho \sim (\pi^{\prime}_{-i}, \pi^t_{i})\)
            \State \quad Train \(\mu_i\) over trajectory \(\rho \sim ( \mu_{i},\pi^{\prime}_{-i})\)
            \State Train distilled \(\pi^{\text{distill}}_i\) for E episodes via Eq.~\eqref{eq:loss0}
            \State \(\Pi^{t+1}_{i} = \Pi^{t}_{i} \cup \{\pi^{\text{distill}}_i\}\)
        \Else
            \For{E episodes}
                \State Sample \(\pi_{-i} \sim \sigma^t_{-i}\)
                \State Train oracle \(\pi^{t+1}_i\) over trajectory \(\rho \sim\)
                \State \( (\pi^{t+1}_i, \pi_{-i})\) with diversity regularization
            \EndFor
            \State \(\Pi^{t+1}_{i} = \Pi^{t}_{i} \cup \{\pi^{t+1}_i\}\)
        \EndIf
    \EndFor
    \State Compute missing entries in \(M^{t+1}\)
    \State Compute a meta-NE \(\sigma^{t+1}\) from \(M^{t+1}\)
\EndFor
\State \textbf{Output:} Current meta-NE for each player
\end{algorithmic}
\end{algorithm}

\section{Experiments}

In this section, we aim to evaluate the effectiveness of Conflux-PSRO in improving BR approximations and reducing exploitability. We compare its performance against existing state-of-the-art PSRO variants, including vanilla PSRO~\cite{lanctot2017unified}, PSRO$_{rN}$\cite{balduzzi2019open}, Pipeline-PSRO~\cite{mcaleer2020pipeline} and PSD-PSRO~\cite{yao2023policy}. We conduct our evaluations on a range of extensive games, including the relatively simple Leduc Poker and more complex games like Goofspiel and Liar's Dice. For Leduc Poker and Goofspiel, we measure and report the exploitability of the meta-NE throughout the training process. Additionally, we perform single-iteration experiments on both Leduc Poker and Goofspiel within a population generated by PSD-PSRO to observe the improvements in BR performance brought by Conflux-PSRO. In the case of Liar's Dice, where calculating exact exploitability is computationally infeasible, we approximate it by training a BR. Lastly, for our ablation studies, we investigate the effects of varying the number of sub-policies, the interval interation for \emph{conflux} applied, and the use of distillation and diversity regularization within the Conflux-PSRO framework.

\label{section-experiment}
\begin{figure*}[h]
    \centering
    \vspace{-10pt}
    \hspace*{0.01\textwidth} 
    \begin{subfigure}[b]{0.45\textwidth}
        \centering
        \includegraphics[width=\linewidth]{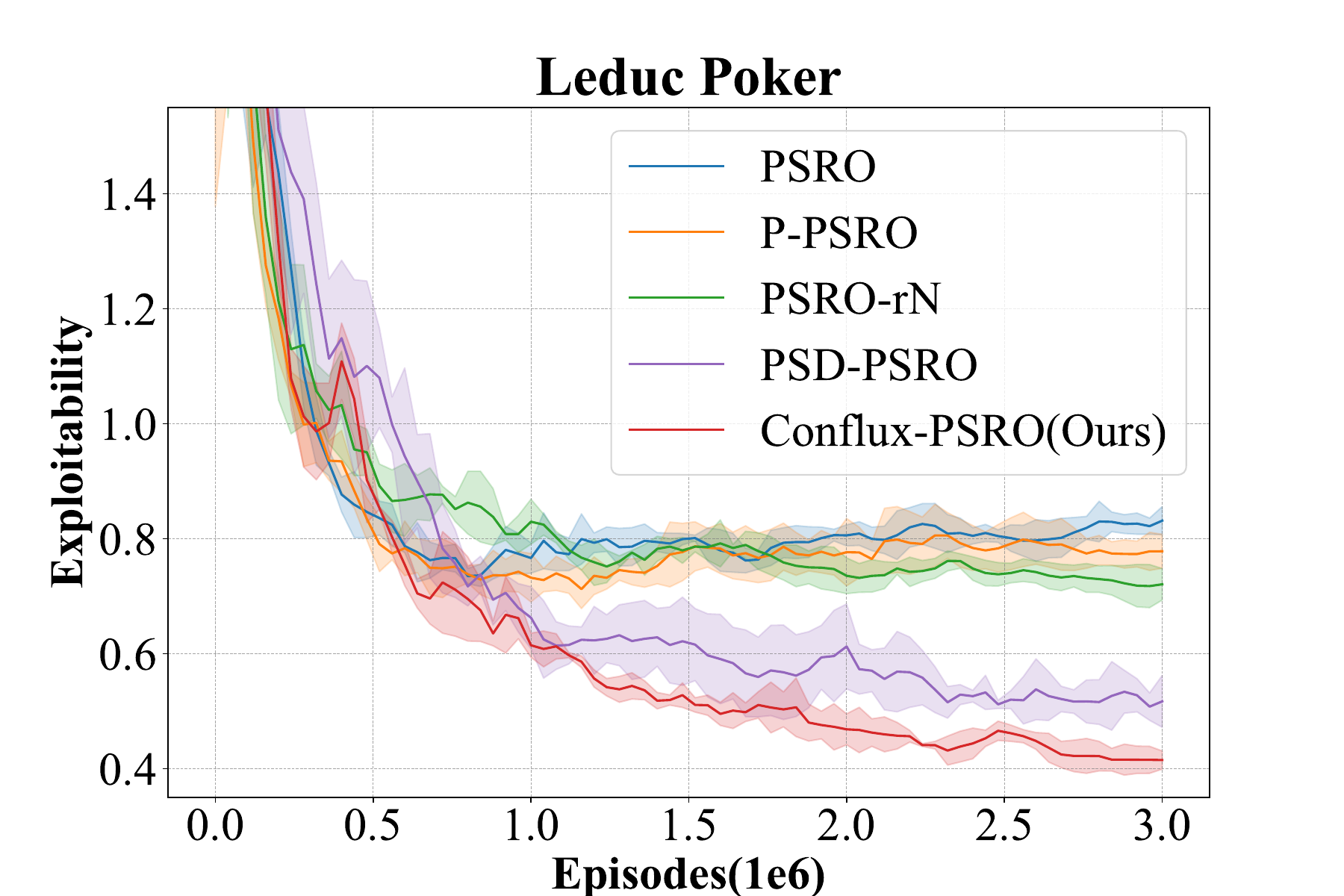}
        \caption{Leduc Poker} 
        \label{leduc_exp}
    \end{subfigure}
    \hspace{0.02\textwidth} 
    \begin{subfigure}[b]{0.45\textwidth}
        \centering
        \includegraphics[width=\linewidth]{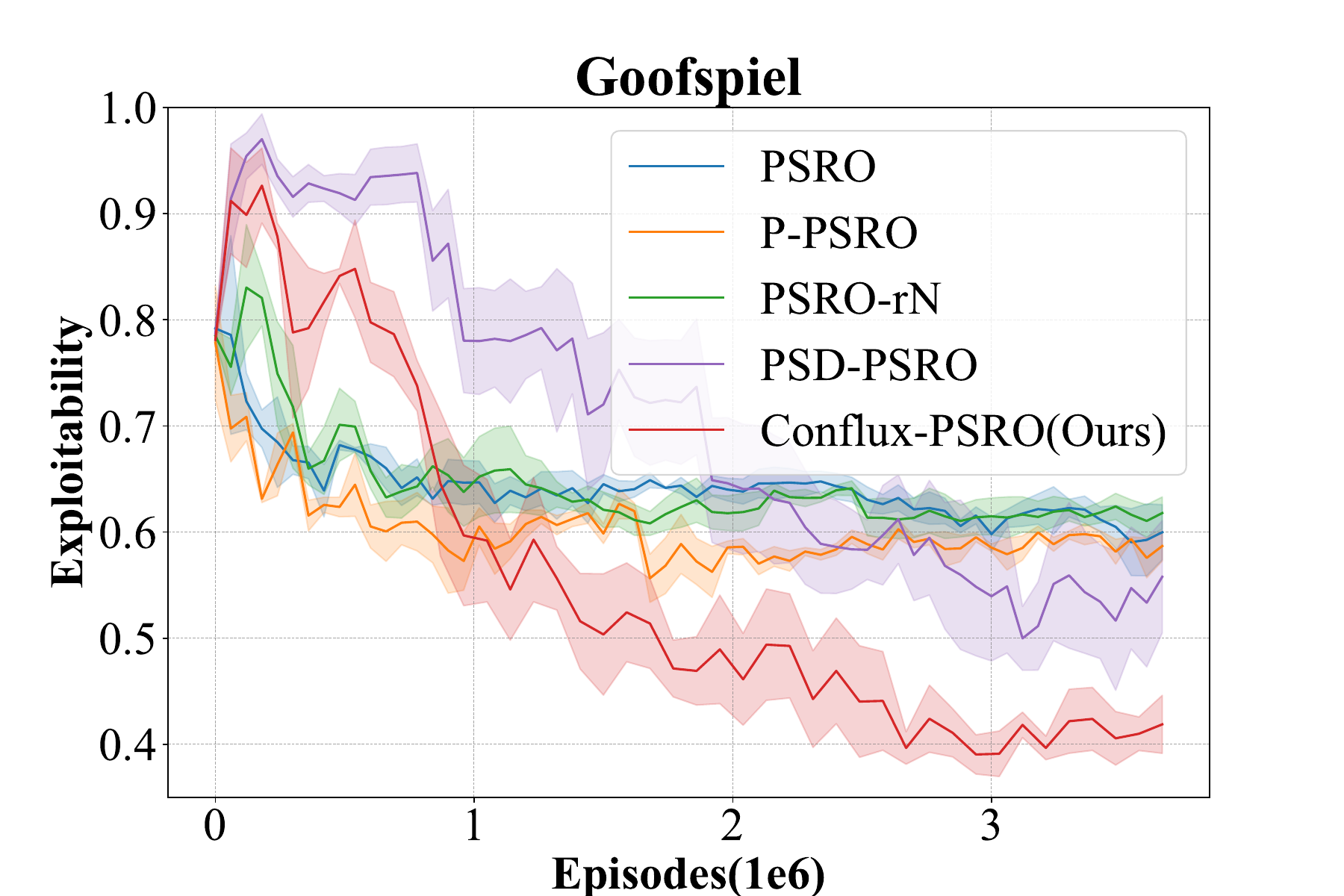}
        \caption{Goofspiel}
        \label{goofspiel_exp}
    \end{subfigure}
    \caption{\emph{Exploitability} on Leduc Poker and Goofspie with 2e5 and 3e5 episodes for training each BR, respectively.} 
    \vspace{-5pt}
\end{figure*}

\begin{figure*}[ht]
    \centering
    \hspace*{0.01\textwidth} 
    \begin{subfigure}[b]{0.45\textwidth}
        \centering
        \includegraphics[width=\linewidth]{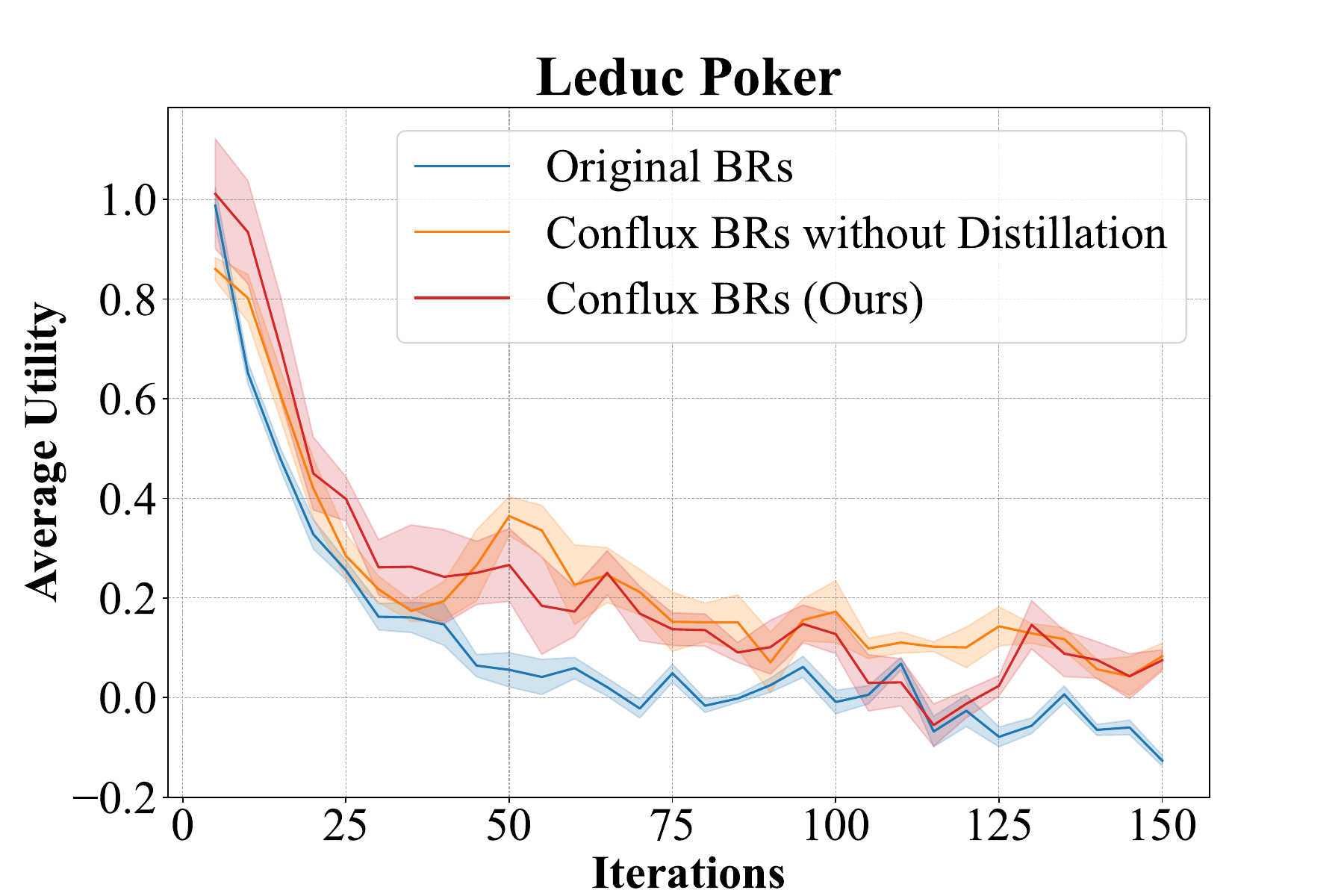}
        \caption{Leduc Poker} 
        \label{leduc_utility}
    \end{subfigure}
    \hspace{0.02\textwidth} 
    \begin{subfigure}[b]{0.45\textwidth}
        \centering
        \includegraphics[width=\linewidth]{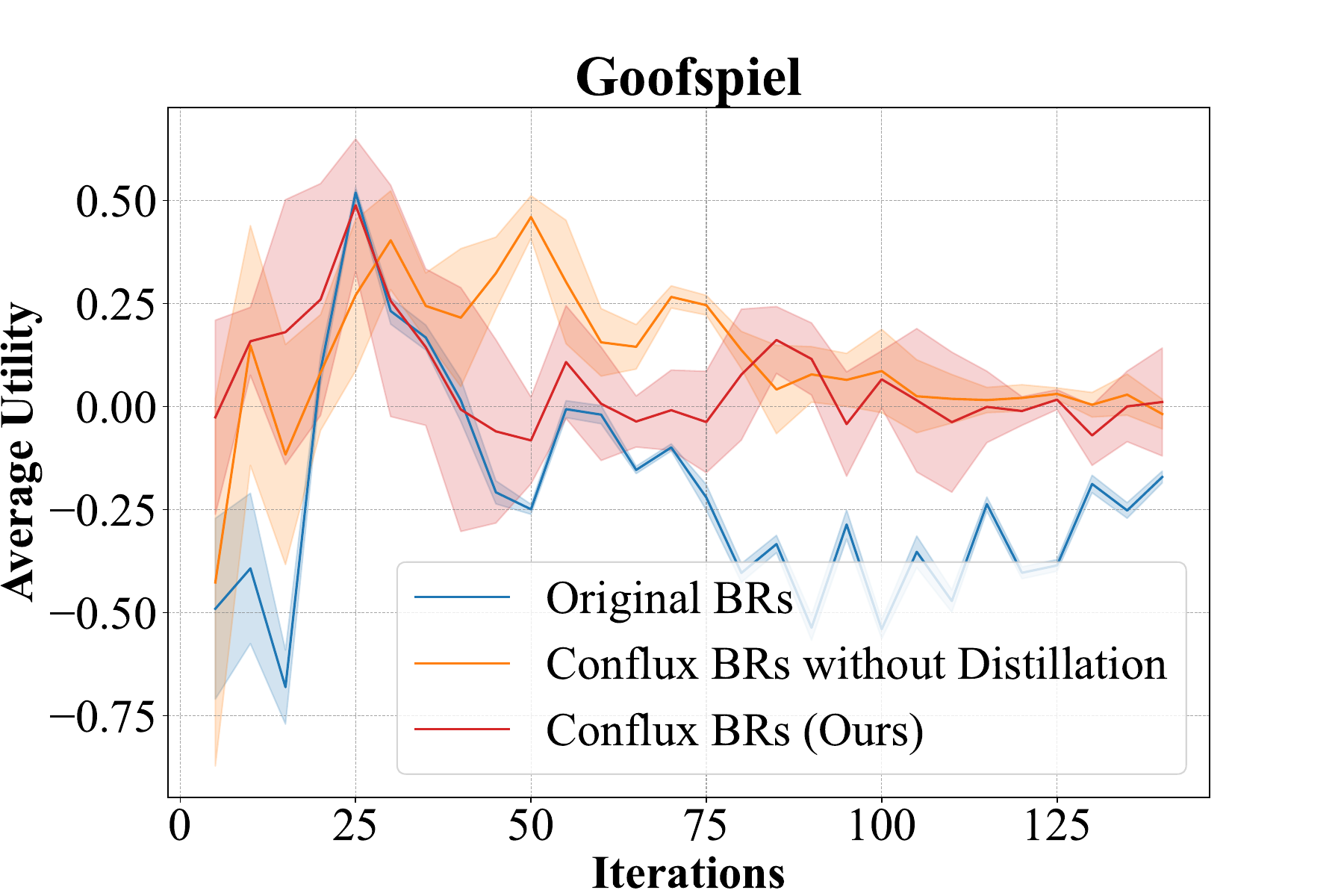}
        \caption{Goofspiel}
        \label{goofspiel_utility}
    \end{subfigure}
    \caption{BR utility generated by PSD-PSRO, Conflux-PSRO, and Conflux-PSRO without distillation after one training iteration.}
\end{figure*}


\subsection{Main Results}

{\bf Leduc Poker.} Leduc Poker is a simplified variant of poker \cite{southey2012bayes}, featuring a deck with two suits and three cards per suit. Each player antes one chip and is dealt a single private card. As shown in Fig.~\ref{leduc_exp}, Conflux-PSRO significantly reduce \emph{exploitability} compared to other methods. To evaluate the improvement Conflux-PSRO offers for BRs, we conducted a single-round experiment depicted in Fig.~\ref{leduc_utility}. Specifically, we generated a population $\Pi_n = \{\pi_1, \dots, \pi_n\}$ using the PSD-PSRO method. For each iteration $\Pi_t$ within this population, we applied Conflux-PSRO for training one additional interation to produce a new policy $\pi_{t+1}^*$, and compared it to the original policy $\pi_{t+1}$. The comparison was made by evaluating the average utility of these policies against $\Pi_t^{-i}$ based on the meta-NE ensemble on 1000 episodes. Although distillation slightly reduced utility, Conflux-PSRO consistently improved BR performance in nearly every iteration. The cumulative advantage gained from this stronger BR made the population more robust to exploitation.

{\bf Goofspiel.} Goofspiel is a large-scale, multi-stage simultaneous move game, implemented in Openspiel~\cite{lanctot2017unified}. As demonstrated in Fig.~\ref{goofspiel_exp}, Conflux-PSRO achieves lower \emph{exploitability} in this highly non-transitive game. As demonstrated in Fig.~\ref{goofspiel_utility}, Conflux-PSRO achieves lower \emph{exploitability} in this highly non-transitive game. Additionally, in the single-iteration experiment shown in Fig.~2, Conflux-PSRO consistently improves BR performance in almost every generation. Notably, the maximum utility in Goofspiel is 1, and Conflux-PSRO achieves an average improvement of 0.2, which is a significant improvement.

\begin{figure*}[ht]
    \centering
    \hspace*{0.01\textwidth} 
    \begin{subfigure}[b]{0.45\textwidth}
        \centering
        \includegraphics[width=\linewidth]{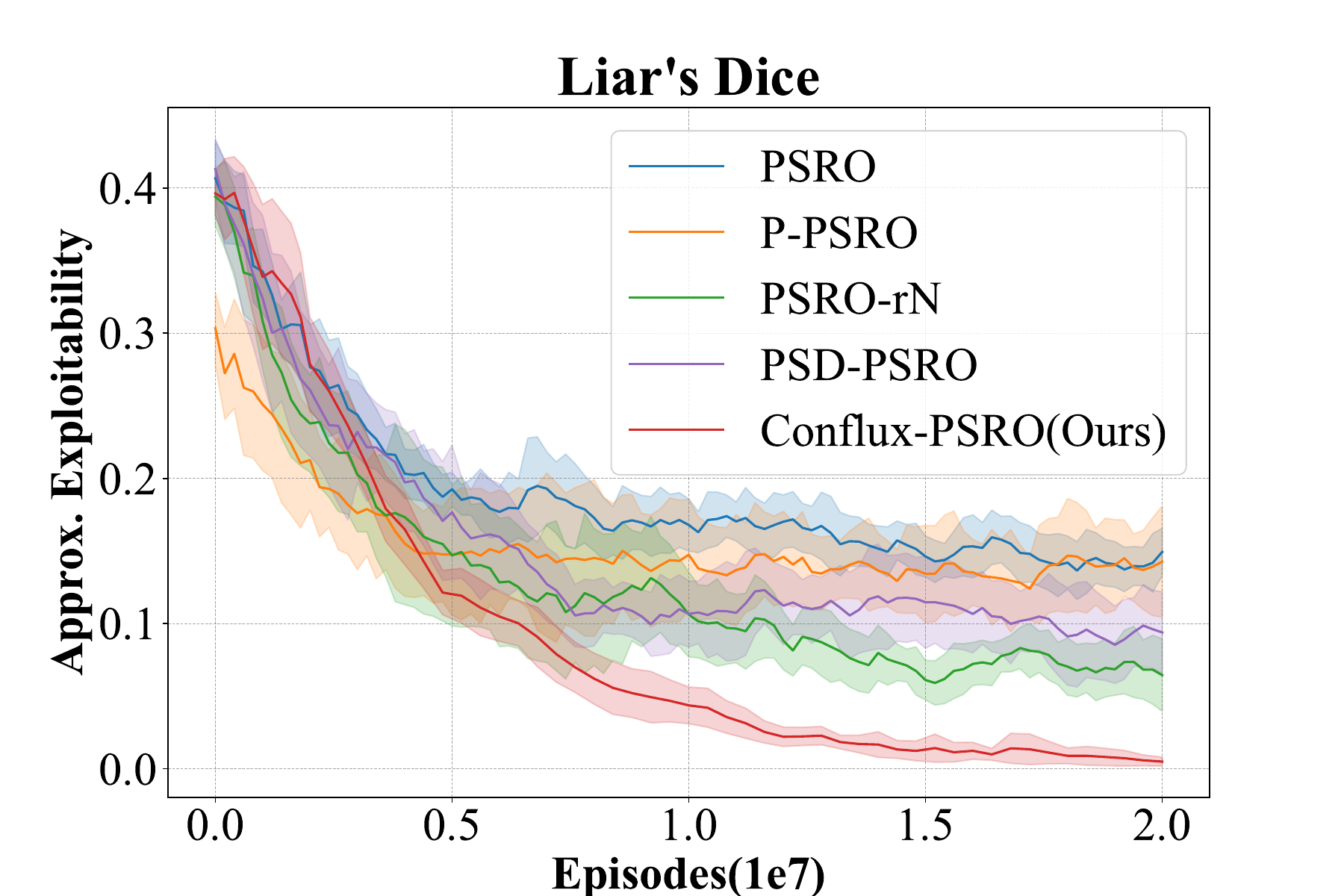}
        \caption{Liars Dice} 
        \label{liarsdice_exp}
    \end{subfigure}
    \hspace{0.02\textwidth} 
    \begin{subfigure}[b]{0.45\textwidth}
        \centering
        \includegraphics[width=\linewidth]{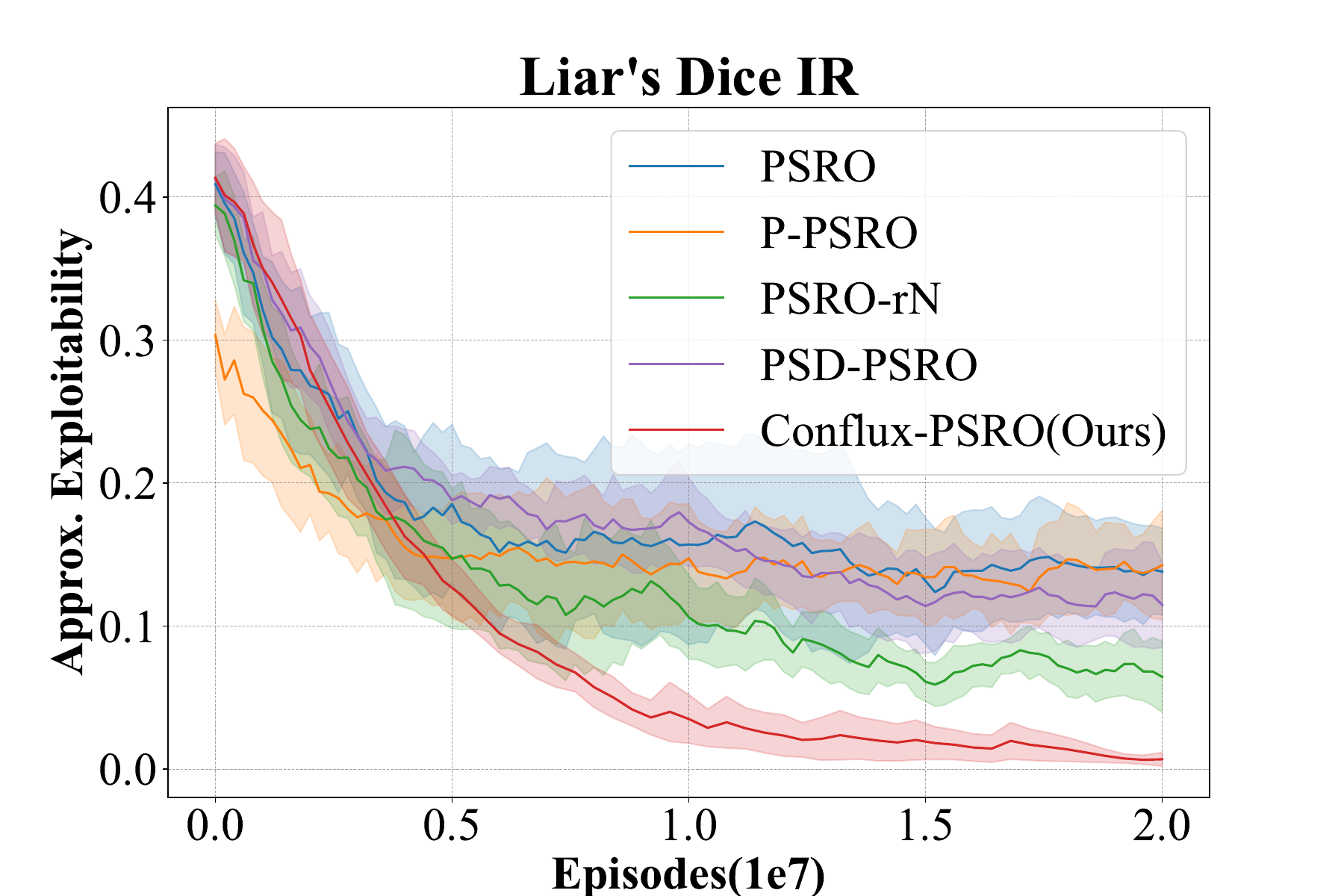}
        \caption{Liars Dice Imperfect Recall}
        \label{liarsdiceir_exp}
    \end{subfigure}
    \caption{\emph{Exploitability} on Liars Dice and Liars Dice IR with 2e5 episodes for training each BR.}
\end{figure*}

{\bf Liar's Dice.} This bluffing game involves players making progressively higher bids about the frequency of a particular die face across all dice in play~\cite{ferguson1991models}. Players alternate between raising bids and challenging the previous bid’s accuracy, risking dice loss (or game defeat) for incorrect challenges or false bids. As illustrated in Fig.~\ref{liarsdice_exp}, Conflux-PSRO achieve lower \emph{approximate exploitability}~\cite{timbers2020approximate} compared to other approaches.

{\bf Liar's Dice Imperfect Recall.} This variant of Liar's Dice incorporates imperfect recall, where players can only remember the current bid, focusing on immediate decision-making. In contrast, perfect recall Liar's Dice allows players to retain a history of bids, which enhances decision-making by considering past actions. This additional memory makes strategic play based on past experiences crucial. As shown in Fig.~\ref{liarsdiceir_exp}, Conflux-PSRO improves PSRO performance, achieving lower \emph{approximate exploitability} in this more complex variant.

\begin{figure*}[ht]
    \centering
    \vspace{-10pt}
    \hspace*{0.01\textwidth} 
    \begin{subfigure}[b]{0.45\textwidth}
        \centering
        \includegraphics[width=\linewidth]{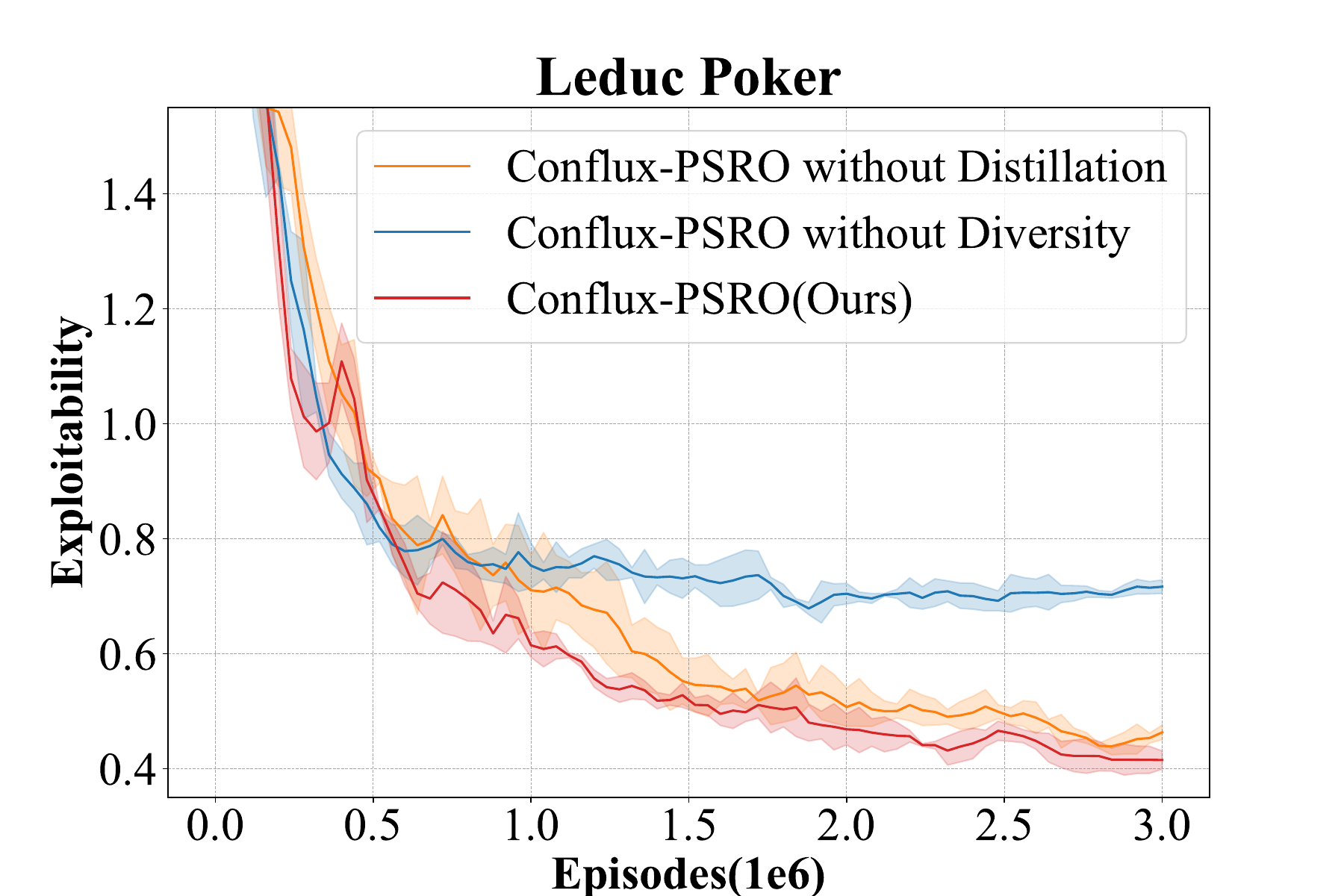}
        \caption{Leduc Poker} 
        \label{leduc_distill}
    \end{subfigure}
    \hspace{0.02\textwidth} 
    \begin{subfigure}[b]{0.45\textwidth}
        \centering
        \includegraphics[width=\linewidth]{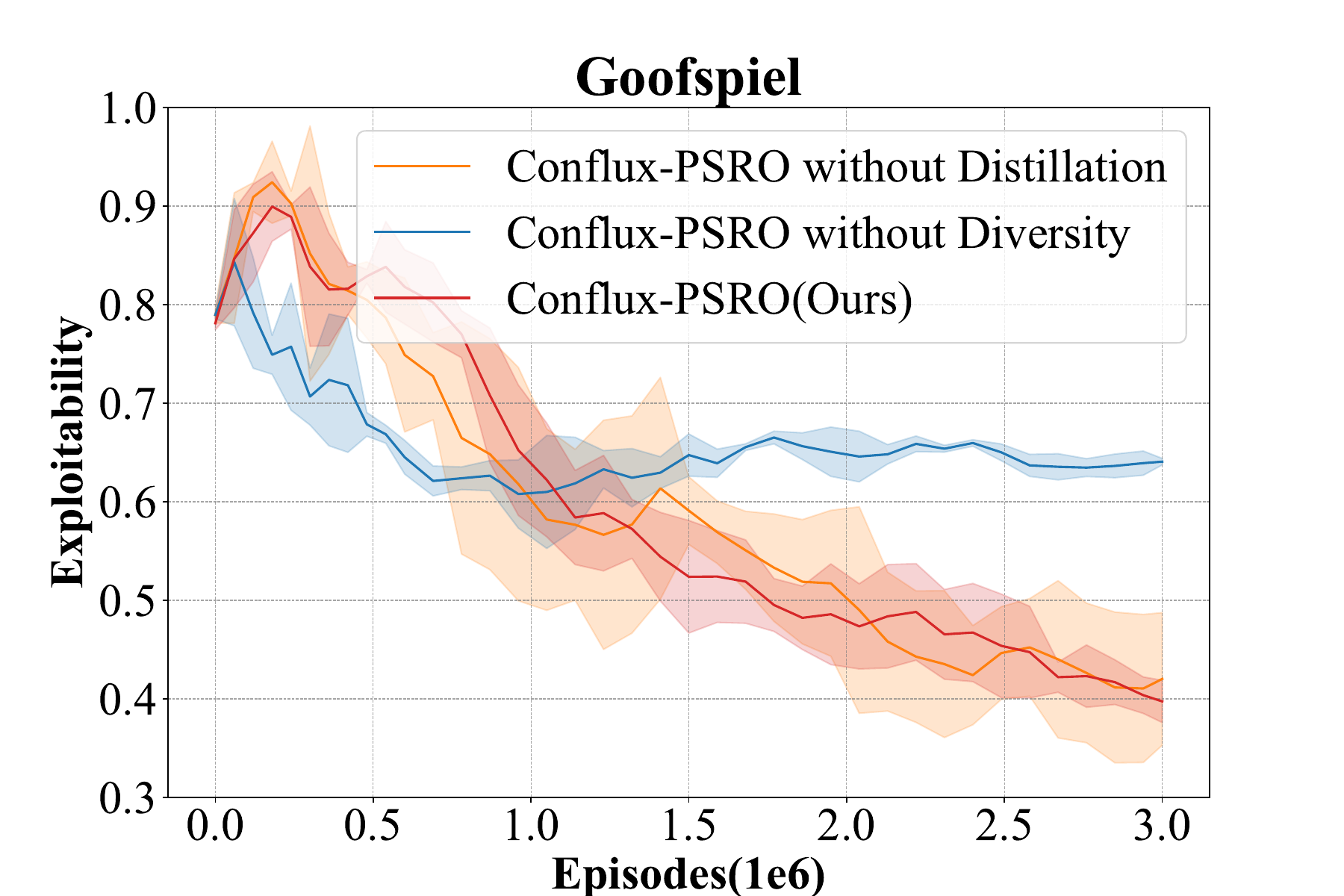}
        \caption{Goofspiel}
        \label{goofspiel_distill}
    \end{subfigure}
    \caption{Ablations on Distillation and Diversity Regularization within Conflux-PSRO.} 
    \vspace{-5pt}
\end{figure*}

\begin{figure}[t]
    \centering
    \begin{subfigure}[b]{0.23\textwidth}
        \centering
        \includegraphics[width=\textwidth]{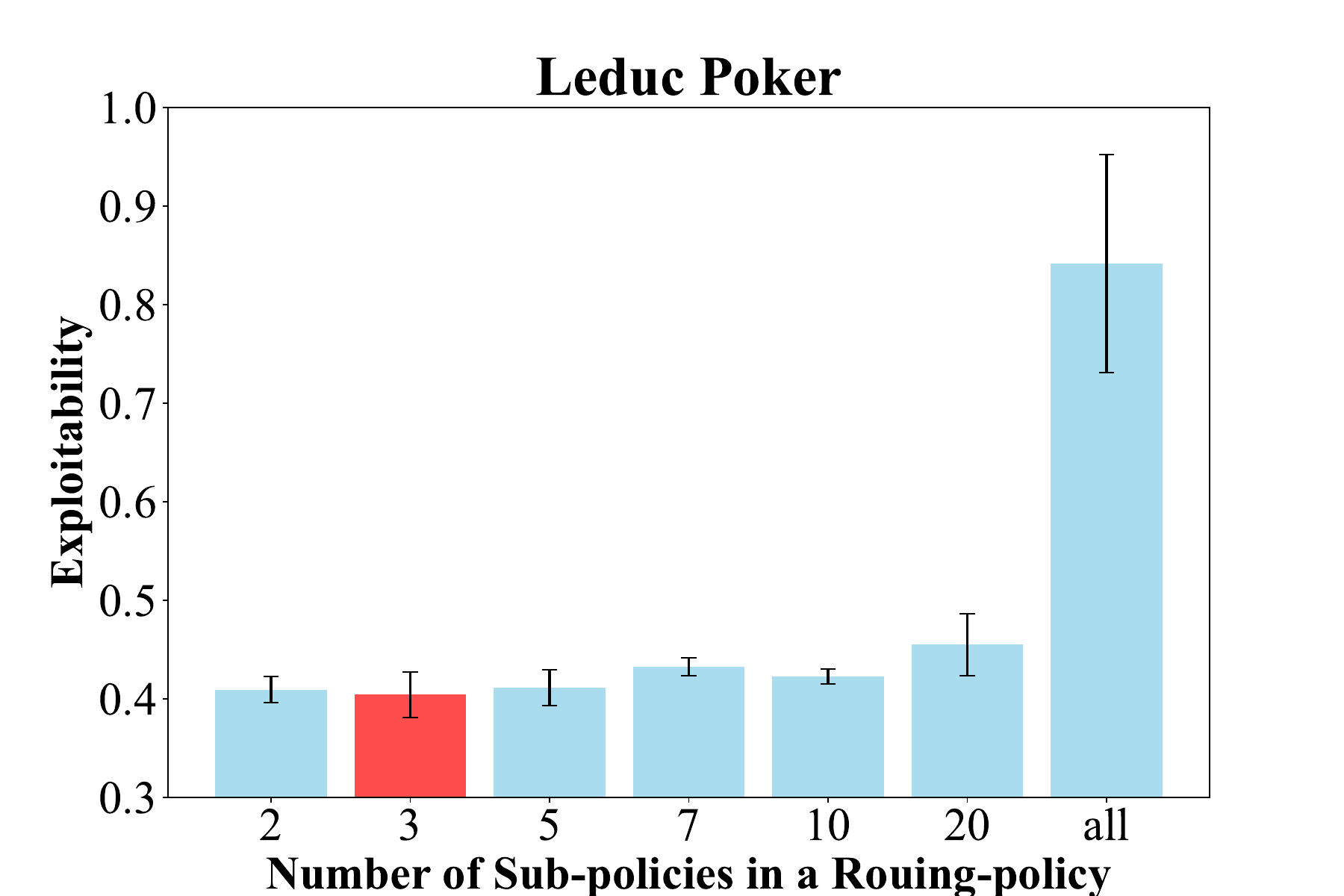}
        \caption{Number of Sub-Policies.}
        \label{leduc_subnum}
        \Description{A plot showing effect of the Number of Sub-Policies on Exploitability in Goofspiel.}
    \end{subfigure}
    \begin{subfigure}[b]{0.23\textwidth}
        \centering
        \includegraphics[width=\textwidth]{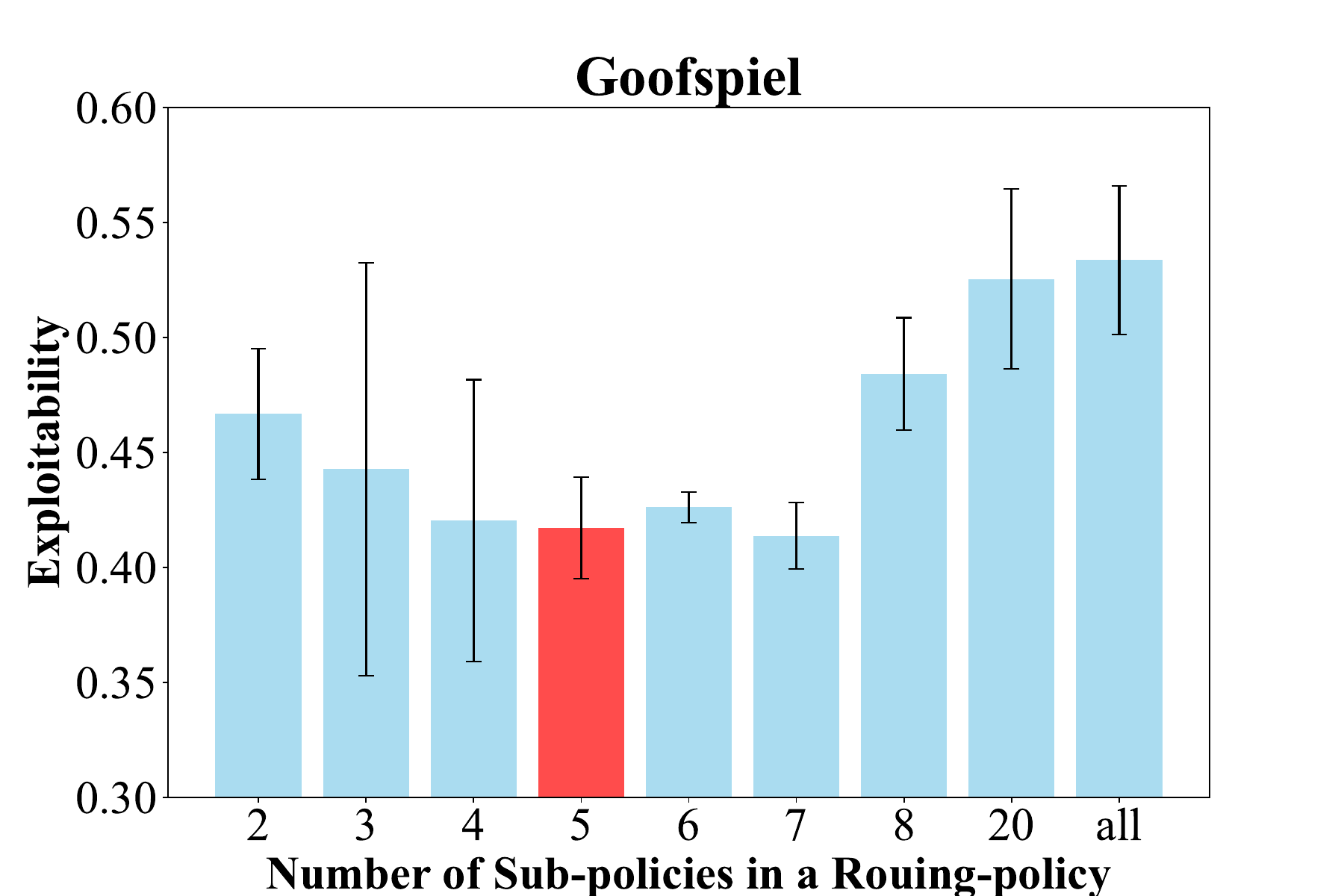}
        \caption{Number of Sub-Policies.}
        \label{goofspiel_subnum}
        \Description{A plot showing effect of the Number of Sub-Policies on Exploitability in Leduc Poker.}
    \end{subfigure}

    \begin{subfigure}[b]{0.23\textwidth}
        \centering
        \includegraphics[width=\textwidth]{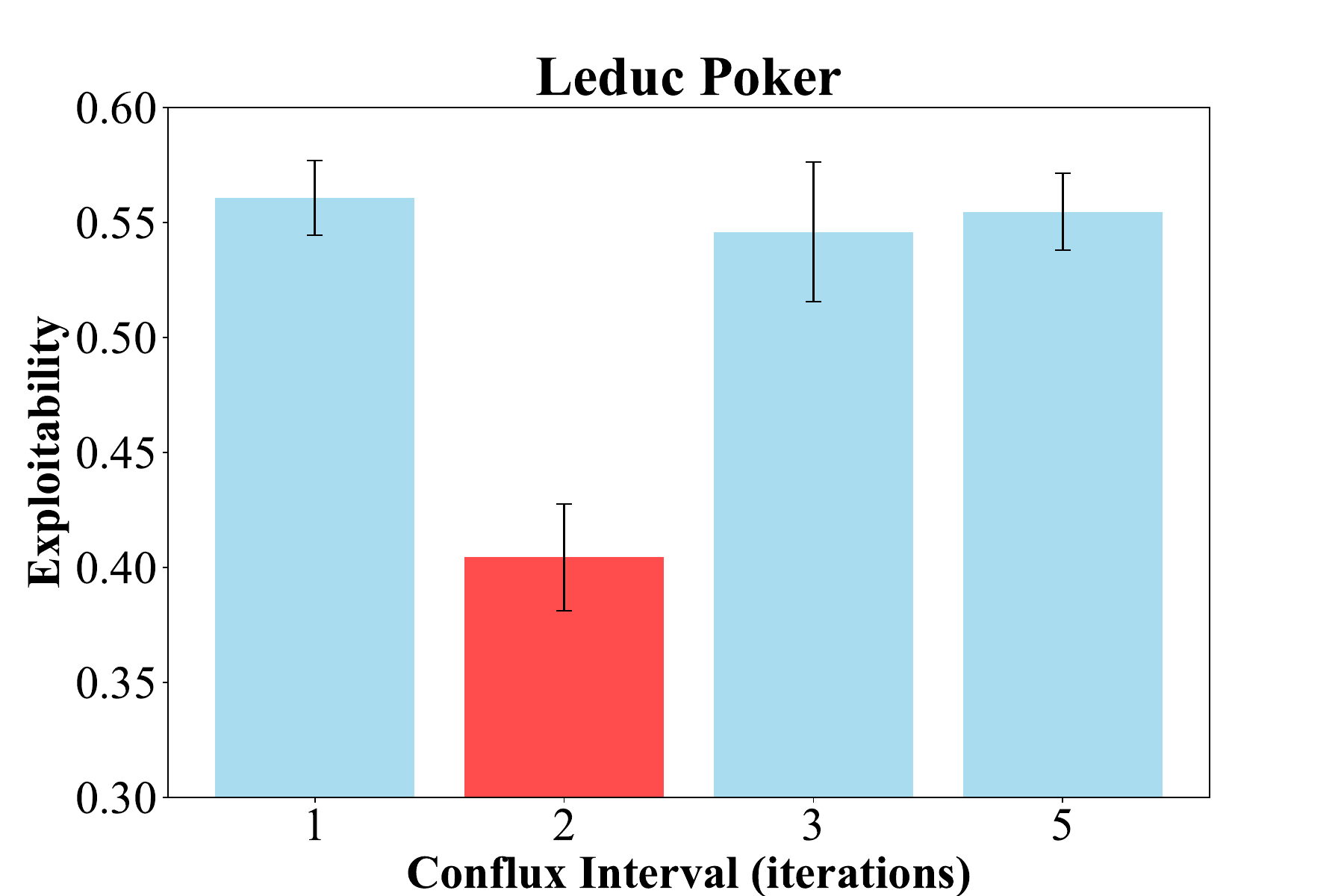}
        \caption{Conflux Interval.}
        \label{leduc_eachiter}
        \Description{A plot showing the Effect of the Interval for Conflux Operation on Exploitability in Goofspiel.}
    \end{subfigure}
    \begin{subfigure}[b]{0.23\textwidth}
        \centering
        \includegraphics[width=\textwidth]{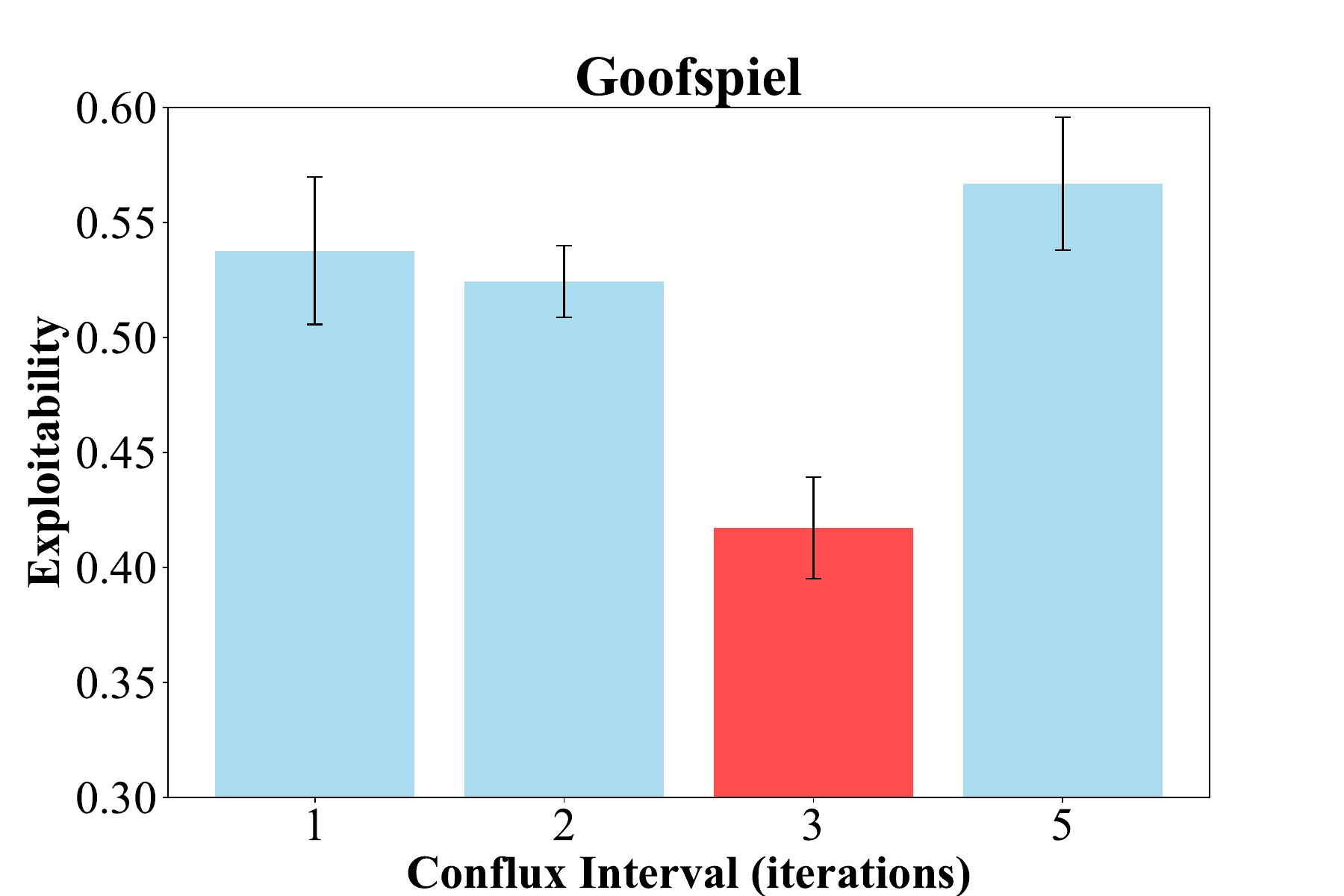}
        \caption{Conflux Interval.}
        \label{goofspiel_eachiter}
        \Description{A plot showing The Effect of the Interval for Conflux Operation on Exploitability in Leduc Poker.}
    \end{subfigure}
    \caption{Ablations on effect of the number of sub-policies and the interval iteration for conflux operation in Leduc Poker and Goofspiel.}
    \Description{Figure contains four subplots of ablations on effect of the number of sub-policies and the interval iteration for conflux operation.}
    \label{compare_reward}
\end{figure}

\subsection{Ablation Study}
{\bf Number of Sub-polices.} As shown in Fig.~\ref{leduc_subnum} and Fig.~\ref{goofspiel_subnum}, we conducted this ablation on both Leduc Poker and Goofspiel, varying the number of sub-policies from 2 to the all population. We observed that the optimal number of sub-policies differs for each environment: 3 in Leduc Poker and 5 in Goofspiel. Deviating too far from the optimal number results in diminished performance. The insight here is that environments with higher complexity and stronger non-transitivity require more sub-policies to ensure adequate coverage. However, excessively increasing the number of sub-policies can raise the training difficulty of the routing-policy, which in turn reduces performance. Given the complexity of different environments, we recommend setting the number of sub-policies not significantly higher than the environment's action space. 
{\bf Interval Iteration for Conflux.} As demonstrated in Fig.~\ref{leduc_eachiter} and Fig.~\ref{goofspiel_eachiter}, 
the optimal number of interval iterations differed between environments, with 2 for Leduc Poker and 3 for Goofspiel. The important observation is that in more complex and non-transitive environments, additional exploration through more frequent iterations is required to effectively adapt policies. The trade-off involves balancing the diversity introduced by the normal BR against the performance enhancements achieved by Conflux-BR.  

{\bf Distillation and Diversity Regularization.} As depicted in Fig.~\ref{leduc_distill} and Fig.~\ref{goofspiel_distill} , the use of distillation in Conflux-PSRO did not significantly reduce the overall performance of the population. Although distillation slightly diminished the performance of the BRs (as seen in Figs. ~\ref{leduc_utility} and Fig.~\ref{goofspiel_utility}), the distilled policies can still be effectively incorporated as a sub-policy into the subsequent routing policy, thereby continuously enhancing the overall capability of the population. Regarding diversity, the lack of diversity exploration considerably reduced the effectiveness of Conflux-PSRO, indicating that maintaining diversity is crucial for the success of the routing policy.

In addition to above parameters, we also examined the starting iteration for the Conflux operation and the size of the selected pool. Empirically, the starting iteration for Conflux is typically set between 10 and 20 iterations, allowing the routing policy to select relatively mature sub-policies of adequate strength to maintain basic performance. For the selected pool size, we set it to 1.5-2 times the number of sub-policies to strike a balance between diversity and performance.


\section{Conclusions and Limitations}

In this paper, we proposed the Conflux-PSRO method, a novel approach designed to generate high-quality BRs. By leveraging the iterative policy generation characteristic of the PSRO algorithm through a MOE structure, our method effectively absorbs and flexibly asign historical policies. By harnessing their combined advantages, we enhance the potential capabilities of this population-based algorithm. Extensive experiments across various non-transitive game environments—including Leduc Poker, Goofspiel, Liar's Dice, and Liar's Dice with Imperfect Recall—demonstrate that Conflux-PSRO enables the population to achieve lower exploitability by the cumulative effect of stronger BRs.

Despite these promising results, our approach has certain limitations. Conflux-PSRO includes several environment-independent hyperparameters that require adjustment. For instance, the number of sub-policies should increase with environmental complexity to ensure sufficient coverage of non-transitive policies. Likewise, the number of interval iterations for the \emph{conflux} operation should also scale with environmental complexity, providing the routing policy with a more diverse set of sub-policies. Additionally, although Conflux-PSRO showed significant improvements in the tested environments, its scalability to larger and more complex games requires further investigation. These limitations could be addressed by exploring adaptive parameter settings, enabling the method to be applied effectively in more complex environments. Furthermore, Conflux-PSRO comprises multiple components, such as the choice of diversity metrics, the distillation method, the design of the routing-policy structure, and the training approach for the routing-policy. Discovering higher-quality and more compatible components in these directions will be of significant importance for enhancing the performance of Conflux-PSRO.



\bibliographystyle{ACM-Reference-Format} 
\bibliography{main}

\clearpage
\newpage 
\appendix
\section{Appendix}
\subsection{Algorithm for Distillation in Conflux-PSRO}
\begin{algorithm}[H]
\caption{Routing-policy Distillation}
\label{alg:distillation}
\begin{algorithmic}[1]
\State \textbf{Input}: routing-policy $\mu_i$, Population $\Pi^t_i$, meta-NE $\sigma^t_{-i}$
\For{many episodes}
    \State Sample opponent policy \(\pi_{-i} \sim \sigma^t_{-i}\)
    \State Collect trajectory \(\tau\) by playing \(\pi_{\text{distill}}\) against \(\pi_{-i}\)
    \State Discount terminal utility $u$ as extrinsic reward \(r_0(s)\) 
    \State Discount $\, -R_{\text{KL}(\pi_i^{\text{distill}} \,||\, \mu_i)}(\tau)$ as imitation reward \(r_1(s)\) 
    \State Discount $\, R_{\text{KL}(\pi_i^{\text{distill}} \,||\, H(\Pi^t)}(\tau)$ as diversity reward \(r_2(s)\) 
    \State Compute total reward \(r(s) = r_0(s) + r_1(s) + r_2(s)\)
    \State Store transition \((s, a, s', r(s))\) in replay buffer
    \State Update \(\pi_i^{\text{distill}}\) using samples from replay buffer via Eq.(~\ref{eq:gradient})
\EndFor
\State \textbf{Output}: Trained distilled policy \(\pi_{\text{distill}}\)
\end{algorithmic}
\end{algorithm}

\section{Benchmark and Implementation Details}
\label{appendix-benchmark}

\begin{table}[htbp!]
\caption{Hyper-parameters for Leduc Poker.}
\centering
\small
\begin{tabular}{|l|l|}
 \hline
 \textbf{Hyperparameters} & \textbf{Value} \\
 \hline
 \multicolumn{2}{|l|}{\textit{Oracle}} \\
 \hline
 Oracle agent & DQN\\
 Replay buffer size & $10^4$ \\
 Mini-batch size & 512\\
 Optimizer & Adam \\
 Learning rate & $5 \times 10^{-3}$\\
 Discount factor ($\gamma$) & 1\\
 Epsilon-greedy Exploration ($\epsilon$)& 0.05\\
 Target network update frequency& 5\\
 Policy network & MLP (256-256-256)\\
 Activation function in MLP & ReLu \\
 \hline
 \multicolumn{2}{|l|}{\textit{PSRO}} \\
 \hline
 Episodes for each BR training & $2 \times 10^4$ \\
 meta-policy solver & Nash \\
 \hline
 \multicolumn{2}{|l|}{\textit{PSD-PSRO}} \\
 \hline
 Episodes for each BR training & $2 \times 10^4$ \\
 Meta-strategy solver & Nash \\
 Diversity weight ($\lambda$) & 1 \\
 \hline
 \multicolumn{2}{|l|}{\textit{Conflux-PSRO}} \\
 \hline
 Conflux start iteration & 10\\
 Conflux interval iteration & 2 \\
 Number of sub-policies & 3 \\
 Number of inferences & 3 \\
 Size of selected pool & 5 \\
 \hline
\end{tabular}
\label{table-leduc}
\end{table}

\begin{table}[htbp!]
\caption{Hyper-parameters for Goofspiel (5 point cards).}
\centering
\small
\begin{tabular}{|l|l|}
 \hline
 \textbf{Hyperparameters} & \textbf{Value} \\
 \hline
 \multicolumn{2}{|l|}{\textit{Oracle}} \\
 \hline
 Oracle agent & DQN\\
 Replay buffer size & $10^4$ \\
 Mini-batch size & 512\\
 Optimizer & Adam \\
 Learning rate & $5 \times 10^{-3}$\\
 Discount factor ($\gamma$) & 1\\
 Epsilon-greedy Exploration ($\epsilon$)& 0.05\\
 Target network update frequency& 5\\
 Policy network & MLP (512-512-512)\\
 Activation function in MLP & ReLu \\
 \hline
 \multicolumn{2}{|l|}{\textit{PSRO}} \\
 \hline
 Episodes for each BR training & $3 \times 10^4$ \\
 meta-policy solver & Nash \\
 \hline
 \multicolumn{2}{|l|}{\textit{PSD-PSRO}} \\
 \hline
 Episodes for each BR training & $3 \times 10^4$ \\
 meta-strategy solver & Nash \\
 diversity weight ($\lambda$) & 1 \\
 \hline
 \multicolumn{2}{|l|}{\textit{Conflux-PSRO}} \\
 \hline
 Conflux start iteration & 20\\
 Conflux interval iteration & 3 \\
 Number of sub-policies & 5 \\
 Number of inferences & 3 \\
 Size of selected pool & 8\\
 \hline
\end{tabular}
\label{table-gf5}
\end{table}

\begin{table}[htbp!]
\caption{Hyper-parameters for Liar's Dice.}
\centering
\small
\begin{tabular}{|l|l|}
 \hline
 \textbf{Hyperparameters} & \textbf{Value} \\
 \hline
 \multicolumn{2}{|l|}{\textit{Oracle}} \\
 \hline
 Oracle agent & Rainbow-DQN\\
 Replay buffer size & $10^5$\\
 Mini-batch size & 512\\
 Optimizer & Adam \\
 Learning rate & $5 \times 10^{-4}$\\
 Learning rate decay&linear decay\\
 Discount factor ($\gamma$) & 0.99\\
 Epsilon-greedy Exploration ($\epsilon$)& 0.05\\
 Target network update frequency& 5\\
  Network soft update ratio&0.005\\
 Prioritized Experience Replay parameter&0.6\\
  Important sampling parameter&0.4\\
 Gradient clip&10\\
 Policy network & MLP (256-256-128)\\
 Activation function in MLP & ReLu \\
 \hline
 \multicolumn{2}{|l|}{\textit{PSRO}} \\
 \hline
 Episodes for each BR training & $2 \times 10^5$\\
 meta-policy solver & Nash \\
 \hline
 \multicolumn{2}{|l|}{\textit{PSD-PSRO}} \\
 \hline
 Episodes for each BR training & $2 \times 10^5$\\
 Meta-strategy solver & Nash \\
 Diversity weight ($\lambda$) & 1 \\
 \hline
 \multicolumn{2}{|l|}{\textit{Conflux-PSRO}} \\
 \hline
 Conflux start iteration & 10\\
 Conflux interval iteration & 2 \\
 Number of sub-policies & 3 \\
 Number of inferences & 3 \\
 Size of selected pool & 5 \\
 \hline
\end{tabular}
\label{table-liarsdice}
\end{table}

\subsection{Leduc Poker}
In Leduc Poker, we use a two-player setup. We apply the PSRO framework with a Meta-Nash solver, employing DQN as the oracle agent. The parameters for the routing policy are consistent with those of the oracle agent, except for the action dimension. The specific hyper-parameters used for this setup are listed in Table~\ref{table-leduc}. 

\subsection{Goofspiel}
In Goofspiel, we use a two-player, 5-card setup. We adopt a descending order, meaning the cards are bid in the sequence 5, 4, 3, 2, 1. Regarding the return, only the win/loss outcome is considered, with 1 for a win and 0 for a loss. We apply the PSRO framework with a Meta-Nash solver, using DQN as the oracle agent. The parameters for the routing policy are consistent with those of the oracle agent, except for the action dimension. Hyper-parameters are shown in Table~\ref{table-gf5}. 

\subsection{Liar's Dice $\&$ Liar's Dice IR}
In both games, we use a two-player, single-die setup. A player loses the game if they make an incorrect challenge or bid. We apply the PSRO framework with a Meta-Nash solver, employing Rainbow-DQN as the oracle agent. The parameters for the routing policy are consistent with those of the oracle agent, except for the action dimension.  The specific hyper-parameters used for this setup are listed  in Table \ref{table-liarsdice}. 


\end{document}